\documentclass[aps,twocolumn,showpacs,preprintnumbers,amsmath,amssymb,superscriptaddress,floatfix,nofootinbib]{revtex4}
\usepackage{graphicx}
\usepackage{hyperref}
\usepackage{amsmath}
\allowdisplaybreaks[1]
\usepackage{amsfonts}
\usepackage{amssymb}
\usepackage{multirow}
\usepackage{makecell}
\usepackage[dvipsnames]{xcolor}
\usepackage{booktabs}
\usepackage{tabularx}
\usepackage{subfigure}
\usepackage{makecell}
\usepackage{ulem}
\usepackage{longtable}

\usepackage{soulpos}
\ulposdef{\hlst}{%
\rlap{\textcolor{yellow}{\rule[-.75ex]{\ulwidth}{2.5ex}}}%
\rule[.45ex]{\ulwidth}{.1ex}%
}

\begin{document}

\title{Exploration of the LHCb $P_c$ states and possible resonances in a unitary coupled-channel model }

\author{Chao-Wei Shen} \email{c.shen@fz-juelich.de}
\affiliation{Institute for Advanced Simulation (IAS-4), \\ Forschungszentrum J\"ulich, D-52425 J\"ulich, Germany}

\author{Deborah R\"onchen} \email{d.roenchen@fz-juelich.de}
\affiliation{Institute for Advanced Simulation (IAS-4), \\ Forschungszentrum J\"ulich, D-52425 J\"ulich, Germany}

\author{Ulf-G. Mei\ss{}ner} \email{meissner@hiskp.uni-bonn.de}
\affiliation{Helmholtz-Institut f\"ur Strahlen- und Kernphysik and Bethe Center for
Theoretical Physics,\\ Universit\"at Bonn, D-53115 Bonn, Germany}
\affiliation{Institute for Advanced Simulation (IAS-4), \\ Forschungszentrum J\"ulich, D-52425 J\"ulich, Germany}
\affiliation{Tbilisi State University, 0186 Tbilisi, Georgia}

\author{Bing-Song Zou} \email{zoubs@mail.tsinghua.edu.cn}
\affiliation{Department of Physics, Tsinghua University, Beijing 100084, China}
\affiliation{CAS Key Laboratory of Theoretical Physics, Institute of Theoretical Physics, \\
Chinese Academy of Sciences, Beijing 100190,China}
\affiliation{School of Physics, University of Chinese Academy of Sciences (UCAS), Beijing 100049, China}

\author{Yu-Fei Wang} \email{wangyufei@ucas.ac.cn}
\affiliation{Institute for Advanced Simulation (IAS-4), \\ Forschungszentrum J\"ulich, D-52425 J\"ulich, Germany}
\affiliation{School of Nuclear Science and Technology, University of Chinese Academy of Sciences (UCAS), Beijing 101408, China}

\date{\today}

\begin{abstract}
We extend our previous study of the interactions of $\bar{D}^{(*)} \Lambda_c - \bar{D}^{(*)}\Sigma_c^{(*)}$ within an analytic, unitary, coupled-channel approach by including the $J/\psi p$ channel and performing fits to the $J/\psi p$ invariant mass distributions in the $\Lambda_b^0 \to J/\psi p K^-$ decay by the LHCb Collaboration.
We take into account the contributions of $t$-channel pseudoscalar and vector meson exchanges and $u$-channel baryon exchanges in this work, and a series of bound states and resonances with different spin-parities are dynamically generated.
According to the results, four states have a significant impact on the physical observables, with two having a spin-parity of $1/2^-$ and the other two having $3/2^-$. 
These states can be associated with the LHCb hidden-charm pentaquarks.
We further provide the coupling strength between each pole and different channels, which provides some insights on how to search for them in future experiments.
Moreover, we also search for poles in higher partial waves up to $J=7/2$ and find indications for the existence of several states with larger widths.
%
\end{abstract}

\maketitle

\section{Introduction}

During the past decade, the LHCb collaboration has reported the existences of several pentaquark candidates in the hidden-charm sector,
starting with  the observation of two  states in the $J/\psi p$ invariant mass distribution in the $\Lambda_b^0 \to J/\psi p K^-$ decay~\cite{LHCb:2015yax}.
One of them is a narrow state located at 4450~MeV, whereas the other state, $P_c(4380)$, has a width of more than 200~MeV.
The latter state does not correspond to a clear peak structure in the observable, but is required for the analysis of the experimental data.
In 2019, the LHCb collaboration updated its results on the $P_c$ states through the collection of a larger amount of data and improvements in data processing~\cite{LHCb:2019kea}.
A narrow pentaquark state $P_c(4312)$ was discovered with a statistical significance of 7.3~$\sigma$, and the observed structure at 4450~MeV has been confirmed to be composed of contributions from the two states $P_c(4440)$ and $P_c(4457)$.
Another pentaquark state $P_c(4337)$, whose significance is in the range of 3.1 to 3.7~$\sigma$, was reported by the LHCb Collaboration in 2021~\cite{LHCb:2021chn}.
The upgrade of the LHCb detector in 2024~\cite{Johnson:2024omq} and the entailed increase of the data base might lead to even more observations of such states.

The narrow width of some of these $P_c$ states hints at the applicability of a molecular interpretation, that is, the pentaquark state is a molecule in which a meson and a baryon are bound.
The existence of such hidden-charm bound state has been predicted a few years before the experimental observation~\cite{Wu:2010jy,Wang:2011rga,Yang:2011wz,Yuan:2012wz,Wu:2012md,Xiao:2013yca,Uchino:2015uha}, and it attracted widespread attention after the experimental discovery, especially in the $\bar{D}^{(*)} \Sigma_c^{(*)}$ molecule picture~\cite{Chen:2015loa,Chen:2015moa,He:2015cea,Burns:2015dwa,Huang:2015uda,Chen:2016heh,Roca:2016tdh,Lu:2016nnt,Shen:2016tzq,Ortega:2016syt,Lin:2017mtz,Liu:2019tjn,Guo:2019kdc,Liu:2019zvb}. 
In Refs.~\cite{Du:2019pij,Xiao:2020frg,Du:2021fmf}, the authors study the hidden-charm pentaquark states in the coupled-channel approach by fitting experimental data, and the results support the nature of these states as $\bar{D}^{(*)} \Sigma_c^{(*)}$ molecules.
The widths and quantum numbers of these states are also given.
In Ref.~\cite{Nakamura:2021qvy}, also based on a fit to experimental data, the double triangle mechanisms with singularities are proposed to reproduce the peak structures of $P_c(4312)$, $P_c(4380)$ and $P_c(4457)$, then only $P_c(4440)$ is a resonance. The role of the triangle diagram mechanism in this
context was first discussed in Ref.~\cite{Guo:2015umn}.
In contrast, Ref.~\cite{Burns:2022uiv} claims that $P_c(4312)$, $P_c(4380)$ and $P_c(4440)$ are resonances with $\bar{D}^{(*)} \Sigma_c^{(*)}$ constituents, while there are various possibilities for the formation of $P_c(4457)$.
In addition to the hadronic molecule interpretation, there are other theoretical hypotheses, based, e.g., on the constituent quark model~\cite{Maiani:2015vwa,Li:2015gta,Ghosh:2015xqp,Wang:2015epa,Xiang:2017byz,Hiyama:2018ukv,Ali:2019npk} or triangle singularities arising from the observed process~\cite{Guo:2015umn,Liu:2015fea,Guo:2016bkl}.
Due to the limited data base all theoretical interpretations of the nature of the states are accompanied by large uncertainties, precluding a firm conclusion so far.
It is also possible that the observed structure is formed by contributions from multiple sources simultaneously.
In combination with a reinforced experimental foundation in the form of an enlarged data base, a more comprehensive theoretical ansatz is still needed to give a reliable description of both the narrow and broad states. To this end, coupled-channel frameworks represent an especially suited tool since the spectrum of possible $P_c$ states can be extracted from a  simultaneous analysis of multiple reaction channels once the corresponding data are available.  
Various coupled-channel methods, which contain different coupled channels, have been used to investigate these $P_c$ states~\cite{Roca:2015dva,Shimizu:2016rrd,Yamaguchi:2016ote,Shimizu:2017xrg,Yamaguchi:2017zmn,Xiao:2019aya,Burns:2019iih,Yalikun:2021bfm}, and the $\bar{D}^{(*)} \Sigma_c^{(*)}$ molecular picture is the most favoured.

The J\"ulich-Bonn (J\"uBo) model is a dynamical coupled-channel (DCC) approach based on the meson-exchange picture that was originally constructed to describe $\pi N$ scattering and extract the light baryon resonance spectrum~\cite{Schutz:1998jx}.
In the last decade it has been expanded to the pion- and photon-induced production of various other final states~\cite{Ronchen:2012eg,Ronchen:2014cna,Ronchen:2022hqk,Wang:2022osj} and to electroproduction reactions~\cite{Mai:2021vsw,Mai:2021aui,Mai:2023cbp}.
In 2017, we first applied this model to explore the possibility of dynamically generated states in the hidden-charm sector~\cite{Shen:2017ayv} in different partial waves, followed by an extended study in Ref.~\cite{Wang:2022oof}. 
The coupled-channel interactions of $\bar{D}^{(*)} \Lambda_c-\bar{D}^{(*)} \Sigma_c^{(*)}$ are studied, and in the energy range of 4.3 to 4.5~GeV, we obtain four narrow $S$-wave states as well as several broad resonances in higher partial waves. 
Based on their pole positions and coupling strengths, we would associate some with  $\bar{D}^{(*)} \Sigma_c^{(*)}$ bound states.

In the present work, the $J/\psi p$ channel, for which experimental data are available, is newly added to the coupled-channel framework, resulting in a total of six coupled channels: $\bar{D}\Lambda_c$, $\bar{D}\Sigma_c$, $\bar{D}^* \Lambda_c$, $\bar{D}^* \Sigma_c$, $\bar{D} \Sigma_c^*$ and $J/\psi p$.
This enables us, for the first time in this approach, to match the free model parameters to the experimental data. Although the connection to experiment is still weak -- data are available only for one of the six channels -- the present study represents an important step towards a comprehensive determination of the baryon spectrum in the hidden-charm sector.
The extension of the model to the $J/\psi p$ channel can also be regarded as preparatory work for the analysis of other reaction channels that may in the future provide complementary information on the $P_c$ states, like $J/\psi p$ photo- and electroproduction, e.g. at GlueX~\cite{GlueX:2019mkq}, LHC~\cite{Goncalves:2019vvo} or the planned Electron-Ion Collider in China (EicC)~\cite{Wang:2023thy}.

This work is organized as follows.
In Sec.~\ref{Sec:theory} we provide a concise theoretical framework for the calculation.
The results obtained by fitting the experimental observables and the corresponding pole positions in various partial waves are presented and discussed in Sec.~\ref{Sec:result}.
Finally, a brief summary is given in Sec.~\ref{Sec:summary}.
%

\section{Theoretical Framework and Numerical Details} \label{Sec:theory}

The scattering equation in J\"uBo DCC model that describes the coupled-channel interactions in a certain partial wave reads:
\begin{align} \label{Eq:LSEq}
   & T_{\mu\nu}(p^{\prime\prime}, p^\prime, z) = V_{\mu\nu}(p^{\prime\prime}, p^\prime, z) \nonumber \\ & + 
        \sum_\kappa \int_0^\infty {\rm d}p\, p^2 V_{\mu\kappa}(p^{\prime\prime}, p, z) 
         G_\kappa(p, z) T_{\kappa\nu}(p, p^\prime, z),
\end{align}
where $z$ is the center-of-mass energy, $p$, $p^{\prime}$ and $p^{\prime\prime}$ are the three-momenta of the intermediate, initial and final states in the center-of-mass frame, which may be on- or off-shell.
The subscripts $\mu$, $\nu$, $\kappa$ are channel indices, which also contain information on spin, orbital angular momentum, total angular momentum and isospin. 
The propagator (two-point function)  $G_\kappa(p, z)$ of the intermediate channel is given by:
\begin{align}
    G_\kappa(p, z) = \frac1{z-E_\kappa(p)-\omega_\kappa(p)+i\varepsilon},
\end{align}
where $E_\kappa(p)$ and $\omega_\kappa(p)$ denote the on-shell energies of the baryon and meson in channel $\kappa$, respectively.
More details on the J\"uBo DCC model, such as the analytic structure and the amplitude decomposition can be found in Refs.~\cite{Doring:2009yv,Ronchen:2012eg} and references therein.

The transition amplitude $V$, which describes the exchange of a meson in a $t$-channel or a baryon in a $u$-channel process and is iterated in Eq.(\ref{Eq:LSEq}) and can be decomposed into:
\begin{equation} \label{Eq:potential}
    V = N F_{1}(q) F_{2}(q)(\mathrm{IF}) \mathcal{V}.
\end{equation}
where $N$ is a kinematic normalization factor~\cite{Ronchen:2012eg} and (IF) is the isospin factor~\cite{Wang:2022oof,Gasparyan:PhDthesis}.
The values of the isospin factors newly involved in this work are given in Table~\ref{Tab:if}.
The off-shell form factors 
\begin{equation} 
F_i(q)=\left( \frac{\Lambda^2-m_{ex}^2}{\Lambda^2+\vec{q}^2} \right)^n
\end{equation}
depend on the exchanged momentum $q$ and the vertex $i$. The cut-offs $\Lambda$ are process-dependent free model parameters that are fitted to data as explained in Sec~\ref{Sec:result}.
A monopole form , i.e. $n=1$, is applied for the $u$-channel baryon exchange processes, while for $t$-channel meson exchange processes, we use the dipole form factor with $n=2$.
The pseudo-potential ${\cal V}$ of Eq.(\ref{Eq:potential}) is constructed from effective interactions in the framework of time-ordered perturbation theory (TOPT).
The involved effective Lagrangians are:
\begin{align}
	{\cal L}_{PPV} =& i \sqrt2 g_{PPV} (P \partial^\mu P - \partial^\mu P P) V_\mu,	\nonumber \\
	{\cal L}_{VVP} =& \frac{g_{VVP}}{m_V} \epsilon_{\mu\nu\alpha\beta} \partial^\mu V^\nu \partial^\alpha V^\beta P, \nonumber \\
	{\cal L}_{VVV} =& i g_{VVV} \langle V^\mu [V^\nu, \partial_\mu V_\nu] \rangle,	\nonumber \\
	{\cal L}_{BBP} =& \frac{g_{BBP}}{m_P} \bar{B} \gamma^\mu \gamma^5 \partial_\mu P B,	\nonumber \\
	{\cal L}_{BBV} =& - g_{BBV} \bar{B} (\gamma^\mu - \frac{\kappa}{2m_B} \sigma^{\mu\nu} \partial_\nu) V_\mu B,	\nonumber \\
	{\cal L}_{BDP} =& - \frac{g_{BDP}}{m_P} (\bar{B} \partial^\mu P D_\mu + \bar{D}_\mu \partial^\mu P B),	\nonumber \\
	{\cal L}_{BDV} =& i \frac{g_{BDV}}{m_V} \big[ \bar{B} \gamma_\mu \gamma^5 D_\nu (\partial^\mu V^\nu - \partial^\nu V^\mu) \nonumber \\
        & + \bar{D}_\mu \gamma_\nu \gamma^5 B (\partial^\mu V^\nu - \partial^\nu V^\mu) \big],	\nonumber \\
	{\cal L}_{DDV} =& g_{DDV} \bar{D}^\tau \left(\gamma^\mu - \frac{\kappa}{2m_D} \sigma^{\mu\nu} \partial_\nu\right) V_\mu D_\tau,
\end{align}
where $P$, $V$, $B$ and $D$ denote the pseudoscalar, vector meson, and baryon octet, respectively, and the decuplet baryons.
There are in total 15 different kinds of exchange pseudo-potentials ${\cal V}$.
The specific expressions for all the amplitudes, the detailed relations between coupling constants, and other relevant theoretical formulae can be found in Ref.~\cite{Wang:2022oof}. 
It should be noted that neither $t$-channel meson  nor $u$-channel baryon exchange processes for $J/\psi N\to J/\psi N$ exist. 

\begin{table}[htbp]
    \centering
    \renewcommand\arraystretch{1.1}
    \caption{The  new isospin factors used in this work. \label{Tab:if}}
    \begin{tabular}{p{3.3cm}<{\centering}|p{1.8cm}<{\centering}|*{2}{p{1.2cm}<{\centering}}}
        \hline
        \hline
        Process & \makecell*{Exchanged\\Particle} & IF$\left(\frac12\right)$ & IF$\left(\frac32\right)$ \\
        \hline
        $\bar{D}^{(*)} \Lambda_c \to J/\psi N$ & $D / D^*$ & 1 & 0 \\
            & $\Lambda_c$ & 1 & 0 \\
        $\bar{D}^{(*)} \Sigma_c^{(*)} \to J/\psi N$ & $D / D^*$ & $-\sqrt{3}$ & 0 \\
            & $\Sigma_c$ & $-\sqrt{3}$ & 0 \\
        \hline
        \hline
    \end{tabular}
\end{table}

As pointed out in Ref.~\cite{Wang:2022oof}, there may be multiple possibilities for a given channel to couple to a certain quantum number $J^P$.
The angular momentum structure of all channels included in the present study is given in Table~\ref{Tab:QuNum}, where the notation for quantum numbers is the common spectroscopic notation, $L_{2I,2J}$, with $L$, $I$ and $J$ denoting the orbital angular momentum, isospin and total angular momentum, respectively.

\begin{table*}[htbp]
    \centering
    \renewcommand\arraystretch{1.12}
    \caption{Angular momentum structure of the coupled channels in $I=\frac12$ up to $J=\frac72$, here $S$ is the total spin. \label{Tab:QuNum}}
    \begin{tabular}{l|*{2}{p{0.8cm}<{\centering}}|*{2}{p{0.8cm}<{\centering}}|*{2}{p{0.8cm}<{\centering}}|*{2}{p{0.8cm}<{\centering}}}
        \hline
        \hline
         \multicolumn{1}{r}{$J^P=$} & $\frac12^-$ & $\frac12^+$ & $\frac32^+$ & $\frac32^-$ & $\frac52^-$ & $\frac52^+$ & $\frac72^+$ & $\frac72^-$  \\
        \hline
        $\bar{D}\Lambda_c, \bar{D}\Sigma_c$ & $S_{11}$ & $P_{11}$ & $P_{13}$ & $D_{13}$ & $D_{15}$ & $F_{15}$ & $F_{17}$ & $G_{17}$ \\
        $\bar{D}^* \Lambda_c \,(1),\, \bar{D}^* \Sigma_c \,(1),\, J/\psi p \,(1)\; (S=1/2)$ & $S_{11}$ & $P_{11}$ & $P_{13}$ & $D_{13}$ & $D_{15}$ & $F_{15}$ & $F_{17}$ & $G_{17}$ \\
        $\bar{D}^* \Lambda_c \,(2),\, \bar{D}^* \Sigma_c \,(2),\, J/\psi p \,(2)\; (S=3/2, |J-L|=1/2)$ & - & $P_{11}$ & $P_{13}$ & $D_{13}$ & $D_{15}$ & $F_{15}$ & $F_{17}$ & $G_{17}$ \\
        $\bar{D}^* \Lambda_c \,(3),\, \bar{D}^* \Sigma_c \,(3),\, J/\psi p \,(3)\; (S=3/2, |J-L|=3/2)$ & $D_{11}$ & - & $F_{13}$ & $S_{13}$ & $G_{15}$ & $P_{15}$ & $H_{17}$ & $D_{17}$ \\
        $\bar{D} \Sigma_c^* \,(1)\; (|J-L|=1/2)$ & - & $P_{11}$ & $P_{13}$ & $D_{13}$ & $D_{15}$ & $F_{15}$ & $F_{17}$ & $G_{17}$ \\
        $\bar{D} \Sigma_c^* \,(2)\; (|J-L|=3/2)$ & $D_{11}$ & - & $F_{13}$ & $S_{13}$ & $G_{15}$ & $P_{15}$ & $H_{17}$ & $D_{17}$ \\
        \hline
        \hline
    \end{tabular}
\end{table*}

\begin{figure}[htbp]
    \centering
    \includegraphics[width=0.8\linewidth]{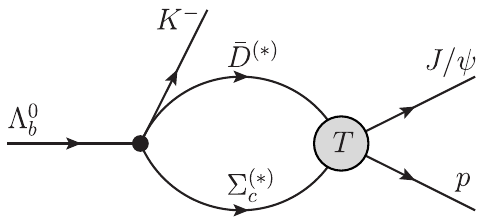}
    \caption{The two-point loop diagram of the $\Lambda_b^0 \to J/\psi p K^-$ decay. \label{Fig:decay} }
\end{figure}

In order to fit the $J/\psi p$ invariant mass distribution of the $\Lambda_b^0 \to J/\psi p K^-$ decay,  we consider the two-point loop diagram as shown in Fig.~\ref{Fig:decay}, where the $P_c$ states can contribute to the final state interactions.
Then for the $J/\psi p$ invariant mass distribution, we have~\cite{Xiao:2016ogq,Xiao:2020frg}:
\begin{equation}
\frac{{\rm d}\Gamma}{{\rm d}m_{J/\psi p}} = \frac1{4(2\pi)^3m_{\Lambda_b}} {\tilde q}_{J/\psi} q_K \left( \sum_{J=1/2}^{5/2}|T^{J}|^2 + |T_{bg}|^2 \right),
\label{eq:inv_mass}
\end{equation}
where $J$ is the total angular momentum of the two-body system and the momenta in the center-of-mass frame are:
\begin{align}
{\tilde q}_{J/\psi} =& \frac{\lambda^{1/2}(m_{J/\psi p}^2,m_{J/\psi}^2,m_p^2)}{2m_{J/\psi p}}, \nonumber \\ 
q_K =& \frac{\lambda^{1/2}(m_{\Lambda_b}^2,m_K^2,m_{J/\psi p}^2)}{2m_{\Lambda_b}},
\end{align}
with $\lambda(\alpha,\beta,\gamma)=\alpha^2+\beta^2+\gamma^2-2\alpha\beta-2\alpha\gamma-2\beta\gamma$  the K\"all$\rm \acute{e}$n function. 
Further, the transition amplitudes of Fig.~\ref{Fig:decay} are given by:
\begin{align}
T^{1/2} =& \sum_{\kappa} g_{\kappa}^{1/2}\;G_{\kappa}\;T^{1/2}_{\kappa \, J/\psi p}, \nonumber \\
T^{3/2} =& \sum_{\kappa} g_{\kappa}^{3/2}\;G_{\kappa}\;T^{3/2}_{\kappa \, J/\psi p}\;q_K, \nonumber \\
T^{5/2} =& \sum_{\kappa} g_{\kappa}^{5/2}\;G_{\kappa}\;T^{5/2}_{\kappa \, J/\psi p}\;q_K^2,
\label{eq:PT}
\end{align}
where $\kappa=\bar{D} \Sigma_c$, $\bar{D}^{*} \Sigma_c$, $\bar{D} \Sigma_c^{*}$ is the channel index of the intermediate loop, $g_{\kappa}^{J}$ are free parameters to be determined from the fit, $T^J_{\kappa \, J/\psi p}$ and $G_\kappa$ are the coupled-channel $T$-matrix and the propagator of Eq.(\ref{Eq:LSEq}).
Note that there is an integration over the momentum of the intermediate state in Eq.(\ref{eq:PT}), which is not written explicitly for simplicity,
and we only consider the total angular momentum of the $J/\psi p$ system up to $J=5/2$. 
According to our test results, 5/2 is sufficient in this analysis, and the contribution of higher angular momenta is completely negligible. 
Moreover, taking higher angular momenta into account will bring more free parameters, which is not supported by the current amount of experimental data.
As in  Refs.~\cite{Xiao:2016ogq,Xiao:2020frg,Wu:2009tu} we include the spectator kaon in Eq.(\ref{eq:PT}) by an additional momentum factor $q_K$ with the correct dependence on the orbital angular momentum in different partial waves (centrifugal barrier).

In addition, the contribution of an incoherent smooth background term is considered in Eq.(\ref{eq:inv_mass}), which compensates for the tree level diagram of the $\Lambda_b^0 \to J/\psi p K^-$ decay and various other contributions which are not explicitly included in the amplitude.
Referring to Ref.~\cite{LHCb:2019kea}, we adopt the polynomial form for this background term as:
\begin{equation}
T_{bg}=a+bs+cs^2,
\label{Eq:bgd}
\end{equation}
which contains the Mandelstam variable $s$ and free parameters $a$, $b$ and $c$, and it does not depend on $J$. 
In Refs.~\cite{Du:2019pij,Xiao:2020frg,Du:2021fmf}, the authors have confirmed that the quality of the fit and the extracted pole results are quite similar regardless of whether such a background contribution contains a resonant term or not.

The fitting of the free parameters of Eqs.(\ref{eq:PT}) and (\ref{Eq:bgd}) to the experimental data in the present work is performed in a $\chi^2$ minimization using MINUIT on the supercomputer JURECA-DC at the J\"ulich Supercomputing Center~\cite{JURECA}.
Here we provide a brief explanation of the free parameters in this work. 
In the J\"uBo model, each exchanged process contains two form factors with a cut-off that is adjusted to experimental data and is not assigned any deep physical meaning, it can be loosely interpreted as accounting for the size of the involved hadrons.
Then in principle, the couplings of the first vertex in the $\Lambda_b^0 \to J/\psi p K^-$ decay, the parameters in the background term $T_{bg}$, and the cut-offs in the form factors should all be free parameters. 
This would amount to a total of 76 parameters, which compared to the total number of experimental data points (175) would result in a highly over-parameterized model.
Therefore, 
we adopt the following strategy to reduce the number of free parameters.
As discussed in Ref.~\cite{Wang:2022oof}, the contributions from $u$-channel diagrams are negligible, and for the $t$-channel diagrams, vector meson exchange is more important than pseudoscalar meson exchange in generating bound states.
In the present study we find that the coupling strengths between the newly added $J/ \psi p$ channel and other anticharmed meson-charmed baryon channels are very small, and the cut-offs related to the $J/ \psi p$ channel have little effect on the data description.
We can confirm that the influence of cut-offs in $u$-channel diagrams can indeed be ignored.
For the $t$-channel processes, especially pseudoscalar meson exchange, we conducted individual tests to understand the importance of each cut-off.
In addition to different initial values, we also tested many different fitting strategies to estimate their impact.
As a result, we set all cut-off values in $u$-channel exchange and some of $t$-channel meson exchanges to fixed values, thereby reducing the number of free parameters in the fitting. 
The values of all involved cut-offs and whether they are fixed or not can be found in Table~\ref{Tab:sum} in Appendix~\ref{App:cutoff}.
%

\section{Results and discussions} \label{Sec:result}

\begin{figure*}[hbpt]
    \centering
    \includegraphics[width=0.44\linewidth]{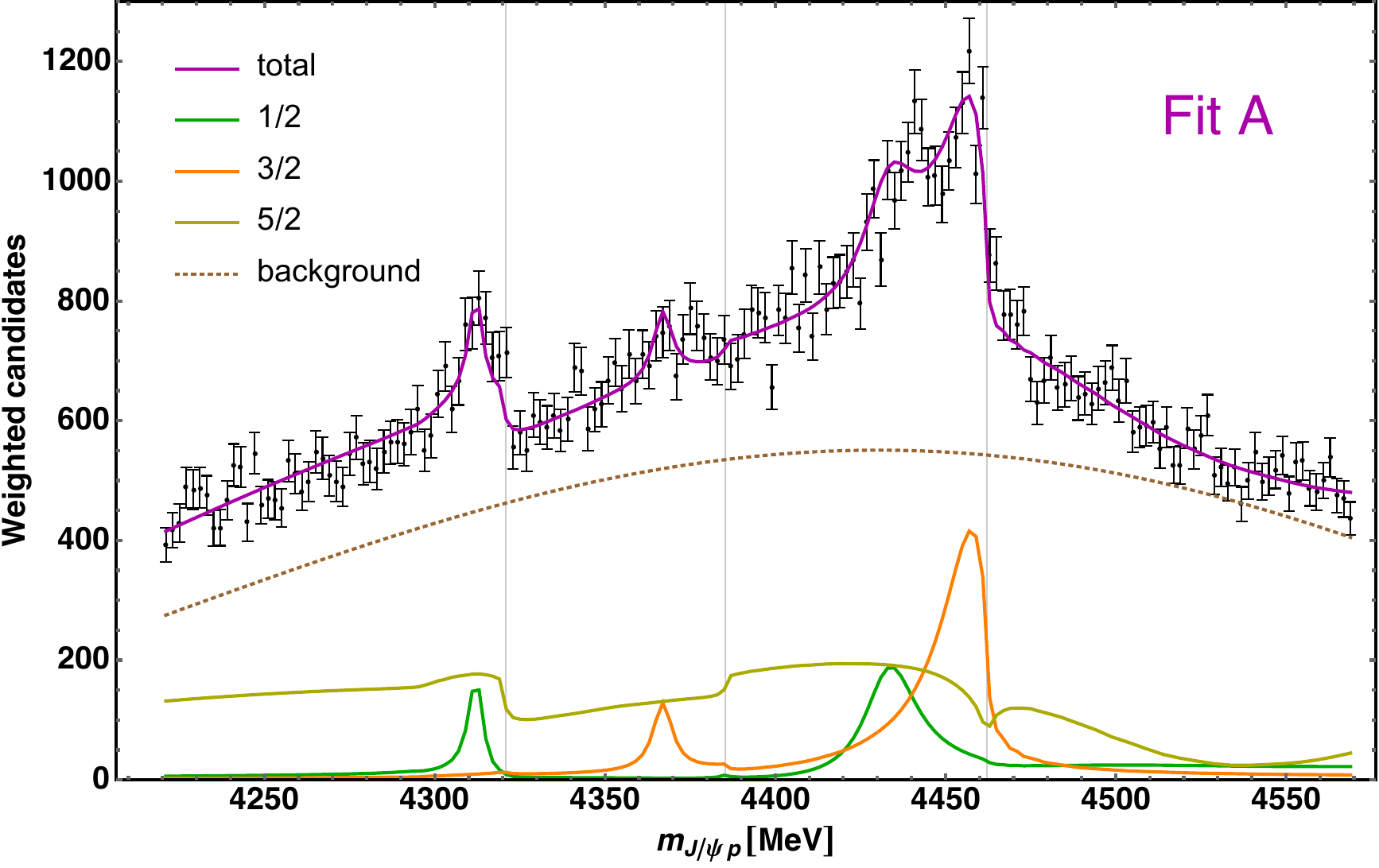} \ \ \
    \includegraphics[width=0.44\linewidth]{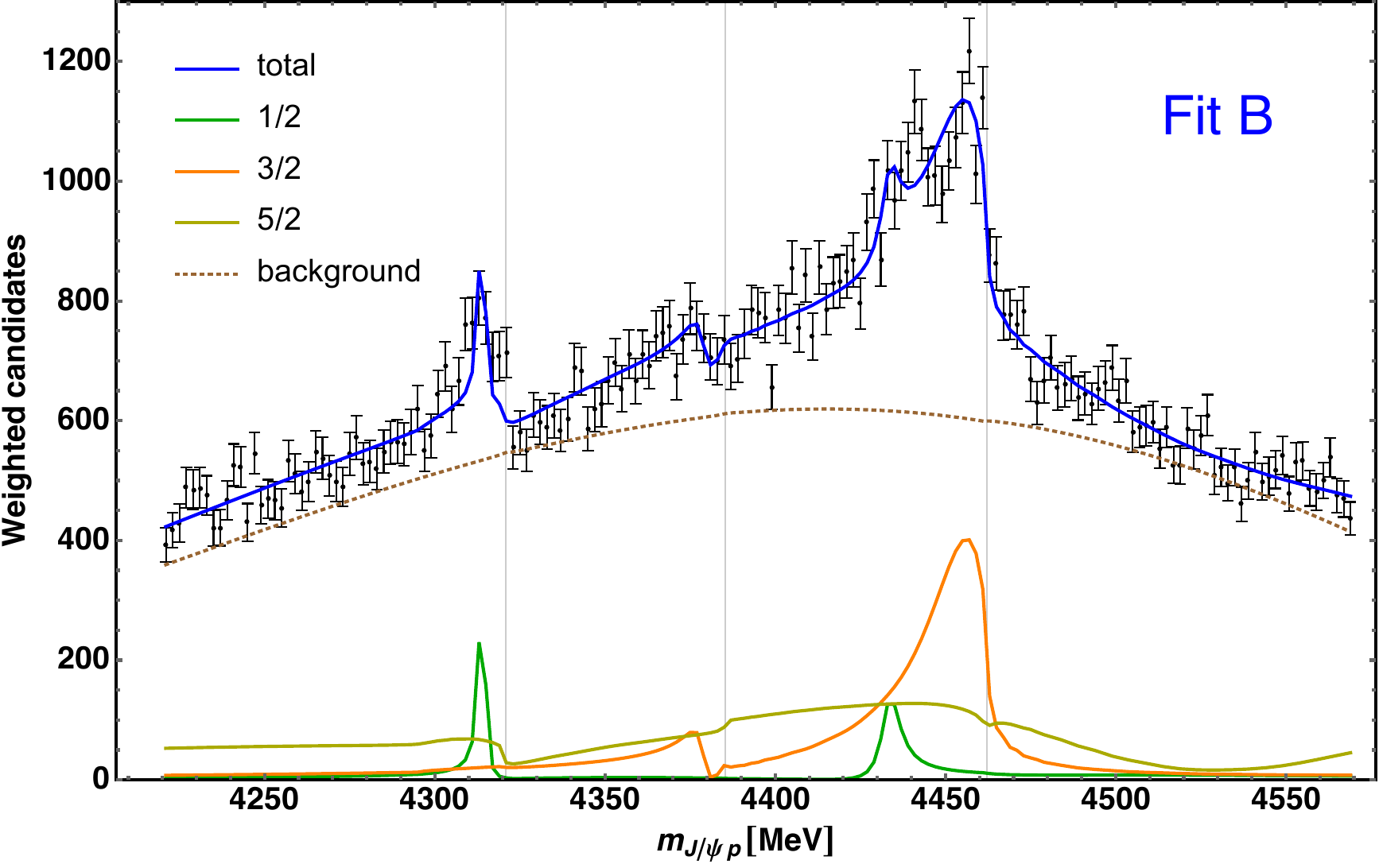} \\
    \includegraphics[width=0.44\linewidth]{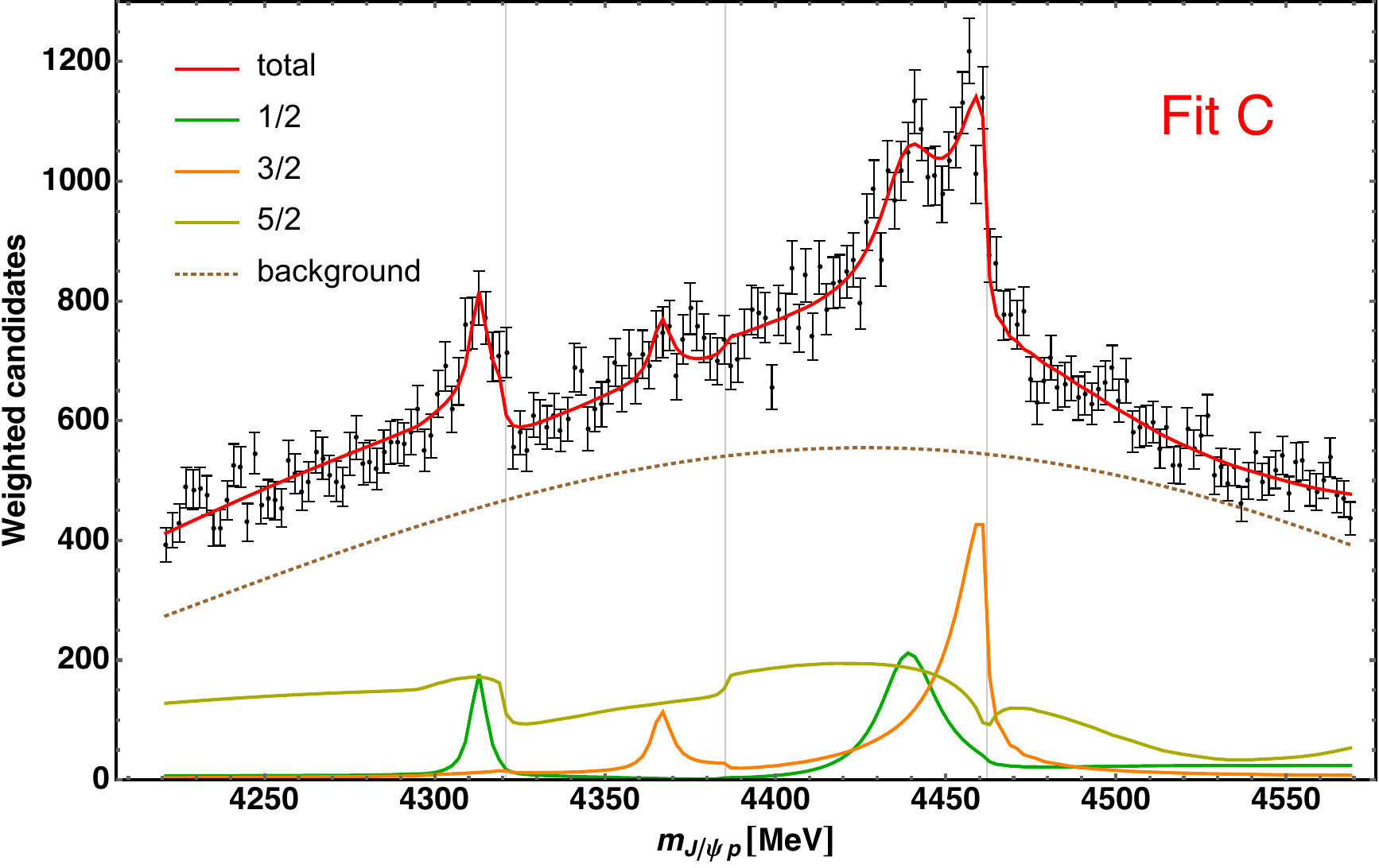} 
    \caption{\label{Fig:fitabc} The three fit results of $J/\psi p$ invariant mass distribution for the $\Lambda_b^0 \to J/\psi p K^-$ decay. Thick solid lines: full results. Green (orange, olive-green) lines: contributions from $J=1/2$ ($3/2$, $5/2$). Dotted lines: background contributions. The three vertical lines are the thresholds of $\bar{D} \Sigma_c$, $\bar{D} \Sigma_c^*$ and $\bar{D}^* \Sigma_c$. }
\end{figure*}

Based on the theoretical formula in the previous section, we are able to fit the $J/\psi p$ invariant mass distribution of the $\Lambda_b^0 \to J/\psi p K^-$ decay by the LHCb Collaboration~\cite{LHCb:2019kea}.
We perform several fits starting from different scenarios in the fit parameter space.
It is found that there are three sets of results, denoted as Fit~A, B and C, describing the data almost equally well, but with different local minima in the parameter space.
Correspondingly, certain differences would appear in the pole positions and residues. 
The fit results of the $J/\psi p$ invariant mass distribution for the decay $\Lambda_b^0 \to J/\psi p K^-$ are shown in Fig.~\ref{Fig:fitabc}.
All three fits are taken into account in the subsequent discussions since the difference in the $\chi^2$ is very small. 
It should be noted that regardless of the efforts to reduce the number of free parameters as explained in the previous section, our model is still over-parameterized due to the limited amount of experimental data. 
The significance of the $\chi^2$ value as a goodness-of-fit criterion is therefore also limited.
The three different fits also allow us to give a rough estimate of the uncertainties of our results.

In all three results in Fig.~\ref{Fig:fitabc}, four clear peak structures can be observed in the $J/\psi p$ invariant mass distribution.
By calculating the contributions of the amplitude $T^J$ with different $J$, which is the quantum number of the $J/\psi p$ system, to the observable, it can be found that two of the four peaks come from $J=1/2$, and the other two are from $J=3/2$.
The contributions of each $J$ and background term are also given in Fig.~\ref{Fig:fitabc}.
The importance of the $J=5/2$ contribution varies among different results, since it can be replaced to some extent by the background term.
An extended data base will likely help to determine the contributions of partial waves with higher angular momentum in the future.

In order to clarify the nature of the four clear peak structures, we extend the $T$ matrix of Eq.(\ref{Eq:LSEq}) to the complex energy plane and search for poles on the second Riemann sheet in various partial waves as discussed in Ref.~\cite{Doring:2009yv}.
We first search for poles that are close to the real axis, i.e. ${\rm Im}(z)<20$~MeV at this point, as their influence is more likely to be reflected in the observables.
There exist two poles in $S_{11}$ and $D_{13}$, respectively.
No other structures are observed close to the real axis in other partial waves, and all the poles obtained in this work are with isospin $1/2$.
The pole positions, quantum numbers and their couplings (residues) to the different channel are listed in Table~\ref{Tab:Spole}.
For details on pole searching and the determination of the coupling strength between a pole and each channel, please refer to Refs.~\cite{Doring:2009yv,Ronchen:2012eg}.
It should be pointed out that the name of the partial wave is always referring to the $\bar{D} \Lambda_c$ channel in this work, and the corresponding spin parity $J^P$ can be found in Table~\ref{Tab:QuNum}.

\begingroup
\renewcommand\arraystretch{1.3}
\LTcapwidth=\textwidth
\begin{longtable*}[ptb]{p{0.5cm}<{\centering}|p{2.8cm}<{\centering}|*5{p{2.7cm}<{\centering}}}
    \caption{Pole positions $z_R~[{\rm MeV}]$, spin-parities $J^P$, and their couplings (residues) to different channels $g~[10^{-3} {\rm MeV}^{-1/2}]$ for the states in $S_{11}$ and $D_{13}$ with $I=\frac12$. For quantum numbers of the channels $(1)$, $(2)$, $(3)$ see Table~\ref{Tab:QuNum}.  \label{Tab:Spole}} \\
    \endfirsthead
    \multicolumn{7}{l}{\tablename\ \thetable{} -- continued from previous page} \\
    \hline
    \endhead
    \hline 
    \multicolumn{7}{r}{{Continued on next page}} \\ 
    \hline
    \endfoot
    \endlastfoot    
        \hline
        \hline
        Fit & $z_R$ , $J^P$ & $g_{\bar{D} \Lambda_c}$ & $g_{\bar{D} \Sigma_c}$ & $g_{\bar{D}^* \Lambda_c(1)}$ & $g_{\bar{D}^* \Lambda_c(2)}$ & $g_{\bar{D}^* \Lambda_c(3)}$ \\
        \hline
        A & $4312.3-i2.4,~\frac12^-$ & $1.1-i0.11$ & $-7.9-i1.1$ & $0.10+i0.33$ & 0 & $-0.024+i0.21$ \\
        B & $4314.0-i0.74,~\frac12^-$ & $0.16-i1.2$ & $-2.8-i2.4$ & $-0.046+i0.16$ & 0 & $-0.074+i0.10$ \\
        C & $4312.9-i2.6,~\frac12^-$ & $1.1+i0.30$ & $-7.8-i1.2$ & $0.091+i0.30$ & 0 & $-0.10+i0.29$ \\
        \cline{3-7}
         & & $g_{\bar{D}^* \Sigma_c(1)}$ & $g_{\bar{D}^* \Sigma_c(2)}$ & $g_{\bar{D}^* \Sigma_c(3)}$ & $g_{\bar{D} \Sigma_c^*(1)}$ & $g_{\bar{D} \Sigma_c^*(2)}$  \\
        \cline{3-7}
         & & $-1.3+i6.7$ & 0 &$2.5+i12$ & $0.080-i0.078$ & $-0.31+i0.88$ \\
         & & $-2.3+i2.6$ & 0 & $-2.4+i3.7$ & $0.074+i0.035$ & $-0.32+i0.22$ \\
         & & $-1.7+i6.3$ & 0 & $1.7+i13$ & $0.082-i0.078$ & $-0.31+i0.73$ \\
        \cline{3-7}
         & & $g_{J/\psi N(1)}$ & $g_{J/\psi N(2)}$ & $g_{J/\psi N(3)}$ & & \\
        \cline{3-7}
         & & $-0.042+i0.0014$ & 0 & $(-1.2+i1.1)\cdot10^{-3}$ & & \\
         & & $(-0.86-i6.0)\cdot10^{-3}$ & 0 & $(-0.67+i1.1)\cdot10^{-3}$ & & \\
         & & $-0.045+i0.0050$ & 0 & $(-1.9+i1.1)\cdot10^{-3}$ & & \\
        \hline
        Fit & $z_R$ , $J^P$ & $g_{\bar{D} \Lambda_c}$ & $g_{\bar{D} \Sigma_c}$ & $g_{\bar{D}^* \Lambda_c(1)}$ & $g_{\bar{D}^* \Lambda_c(2)}$ & $g_{\bar{D}^* \Lambda_c(3)}$ \\
        \hline
        A & $4366.5-i4.7,~\frac32^-$ & $0.25-i0.0024$ & $0.41+i0.019$ & $-0.045-i0.66$ & $-0.078-i0.59$ & $0.73+i1.3$ \\
        B & $4378.7-i4.1,~\frac32^-$ & $0.12+i0.55$ & $0.060+i0.058$ & $-0.14-i0.049$ & $-0.12-i0.041$ & $0.24-i0.047$ \\
        C & $4365.9-i4.7,~\frac32^-$ & $0.25+i0.019$ & $0.49-i0.0019$ & $-0.084-i0.73$ & $-0.12-i0.66$ & $0.88+i1.4$ \\
        \cline{3-7}
         & & $g_{\bar{D}^* \Sigma_c(1)}$ & $g_{\bar{D}^* \Sigma_c(2)}$ & $g_{\bar{D}^* \Sigma_c(3)}$ & $g_{\bar{D} \Sigma_c^*(1)}$ & $g_{\bar{D} \Sigma_c^*(2)}$  \\
        \cline{3-7}
         & & $-1.6+i3.3$ & $1.3+i4.3$ & $1.1-i5.8$ & $0.15-i0.0037$ & $11+i2.1$ \\
         & & $0.16+i0.31$ & $1.0+i0.028$ & $-0.19-i0.62$ & $(7.0+i5.9)\cdot10^{-3}$ & $0.82-i1.2$ \\
         & & $-1.6+i3.9$ & $1.7+i4.6$ & $0.97-i6.2$ & $0.17-i0.027$ & $13+i1.8$ \\
        \cline{3-7}
         & & $g_{J/\psi N(1)}$ & $g_{J/\psi N(2)}$ & $g_{J/\psi N(3)}$ & & \\
        \cline{3-7}
         & & $(0.68-i4.0)\cdot10^{-3}$ & $-0.0031+i0.011$ & $-0.054+i0.19$ & & \\
         & & $(-5.0-i4.2)\cdot10^{-4}$ & $(1.5+i1.2)\cdot10^{-3}$ & $0.024+i0.021$ & & \\
         & & $(0.56-i4.3)\cdot10^{-3}$ & $-0.0030+i0.012$ & $-0.051+i0.20$ & & \\
        \hline
        Fit & $z_R$ , $J^P$ & $g_{\bar{D} \Lambda_c}$ & $g_{\bar{D} \Sigma_c}$ & $g_{\bar{D}^* \Lambda_c(1)}$ & $g_{\bar{D}^* \Lambda_c(2)}$ & $g_{\bar{D}^* \Lambda_c(3)}$ \\
        \hline
        A & $4433.0-i9.3,~\frac12^-$ & $1.7-i0.36$ & $0.65+i0.10$ & $0.19+i0.38$ & 0 & $0.013+i0.0074$ \\
        B & $4433.4-i3.2,~\frac12^-$ & $0.53-i0.94$ & $0.31-i0.074$ & $0.14+i0.32$ & 0 & $0.018-i0.064$ \\
        C & $4438.6-i10,~\frac12^-$ & $1.8+i0.11$ & $0.71+i0.19$ & $0.17+i0.39$ & 0 & $0.016+i0.019$ \\
        \cline{3-7}
         & & $g_{\bar{D}^* \Sigma_c(1)}$ & $g_{\bar{D}^* \Sigma_c(2)}$ & $g_{\bar{D}^* \Sigma_c(3)}$ & $g_{\bar{D} \Sigma_c^*(1)}$ & $g_{\bar{D} \Sigma_c^*(2)}$  \\
        \cline{3-7}
         & & $-1.5+i13$ & 0 & $0.22-i0.69$ & $(-1.1+i0.41)\cdot10^{-3}$ & $0.086-i0.064$ \\
         & & $-2.0+i12$ & 0 & $0.23-i0.63$ & $(-3.5+i1.9)\cdot10^{-3}$ & $0.070-i0.042$ \\
         & & $-1.9+i12$ & 0 & $0.24-i0.51$ & $(-1.1+i0.48)\cdot10^{-3}$ & $0.066-i0.023$ \\
        \cline{3-7}
         & & $g_{J/\psi N(1)}$ & $g_{J/\psi N(2)}$ & $g_{J/\psi N(3)}$ & & \\
        \cline{3-7}
         & & $-0.023+i0.23$ & 0 & $0.0025+i0.044$ & & \\
         & & $-0.032+i0.22$ & 0 & $-0.0033+i0.041$ & & \\
         & & $-0.030+i0.22$ & 0 & $0.0016+i0.043$ & & \\
        \hline
        Fit & $z_R$ , $J^P$ & $g_{\bar{D} \Lambda_c}$ & $g_{\bar{D} \Sigma_c}$ & $g_{\bar{D}^* \Lambda_c(1)}$ & $g_{\bar{D}^* \Lambda_c(2)}$ & $g_{\bar{D}^* \Lambda_c(3)}$ \\
        \hline
        A & $4461.4-i9.7,~\frac32^-$ & $0.85+i0.13$ & $0.67+i0.084$ & $0.27-i0.41$ & $0.20-i0.44$ & $0.27-i0.12$ \\
        B & $4461.5-i14,~\frac32^-$ & $0.86+i0.0095$ & $0.76+i0.033$ & $0.21-i0.60$ & $0.16-i0.66$ & $0.40-i0.024$ \\
        C & $4464.9-i6.8,~\frac32^-$ & $0.79+i0.30$ & $0.64+i0.17$ & $0.29-i0.33$ & $0.22-i0.39$ & $0.26-i0.11$ \\
        \cline{3-7}
         & & $g_{\bar{D}^* \Sigma_c(1)}$ & $g_{\bar{D}^* \Sigma_c(2)}$ & $g_{\bar{D}^* \Sigma_c(3)}$ & $g_{\bar{D} \Sigma_c^*(1)}$ & $g_{\bar{D} \Sigma_c^*(2)}$  \\
        \cline{3-7}
         & & $0.016-i0.055$ & $-0.14+i0.062$ & $-4.1+i6.4$ & $0.077+i0.23$ & $0.039-i2.4$ \\
         & & $-0.041-i0.016$ & $-0.22+i0.0024$ & $-5.2+i7.0$ & $0.053+i0.17$ & $-0.64-i2.9$ \\
         & & $0.036-i0.030$ & $-0.14+i0.015$ & $-4.2+i5.2$ & $0.069+i0.25$ & $0.25-i2.3$ \\
        \cline{3-7}
         & & $g_{J/\psi N(1)}$ & $g_{J/\psi N(2)}$ & $g_{J/\psi N(3)}$ & & \\
        \cline{3-7}
         & & $(-3.9+i7.2)\cdot10^{-3}$ & $0.0093-i0.021$ & $0.13-i0.28$ & & \\
         & & $(-4.1+i7.5)\cdot10^{-3}$ & $0.0097-i0.022$ & $0.14-i0.30$ & & \\
         & & $(-4.5+i6.4)\cdot10^{-3}$ & $0.011-i0.018$ & $0.15-i0.25$ & & \\
        \hline
        \hline
\end{longtable*}
\endgroup

The four poles of Table~\ref{Tab:Spole} are found in all three fit results, and their positions are relatively stable with not too much variation. 
In addition, the relative coupling strength of each channel in the three fits is roughly consistent. 
The first pole at 4313~MeV has a spin-parity of $1/2^-$. It is just below the threshold of $\bar{D} \Sigma_c$ and has a strong coupling to this channel. This is a $\bar{D} \Sigma_c$ bound state. 
Meanwhile, because of its mass, we would also assign this pole to the $P_c(4312)$ state. 
Note that the experimental results of LHCb in 2019 did not explicitly provide spin and parity for the three $P_c$ states.
For the second pole, there are two slightly different position, one at 4366~MeV (Fit A, C) and the other at 4378~MeV (Fit B), indicating that more accurate data are required to give a better description of this energy region and fix the mass of this state.
Based on its position and coupling, we claim that there exists a narrow $\bar{D} \Sigma_c^*$ bound state with $J^P=3/2^-$.
The dominant channel of the other two poles are both $\bar{D}^* \Sigma_c$, with one having the spin-parity of $1/2^-$ and the other being $3/2^-$, indicating that both states can be in $S$-wave.
We assign these two poles to the experimental observed $P_c(4440)$ and $P_c(4457)$ states.
In all our fit results, the third pole is always located 25 to 30~MeV below the threshold of $\bar{D}^* \Sigma_c$, that is, it is likely to be a $\bar{D}^* \Sigma_c$ bound state.
The position of the last pole is also below the threshold of $\bar{D}^* \Sigma_c$ in Fit~A and B, but in Fit~C it is 2.8~MeV above the threshold.
This impedes a classification as a bound state or resonance. 
If this state is, as some people believe~\cite{Guo:2019kdc,Liu:2019zvb,Du:2019pij,Xiao:2020frg,Du:2021fmf,Xiao:2019aya,Yalikun:2021bfm}, a $\bar{D}^* \Sigma_c$ bound state, the result we obtain indicates that its binding energy would be very small. Given the current data situation, our model cannot discriminate between a bound state and a resonance. 
A larger amount of experimental data is required for our model to obtain more accurate results.
In Ref.~\cite{Wang:2023snv}, the authors have studied the composition of $N^*$ or $\Delta$ resonances in three different methods.
This may also be used similarly in the hidden-charm sector to explore the composition of these generated states.

Then, we shift our focus to areas in the complex plane that are not so close to the real axis.
In the J\"uBo model, a notable feature is the ability to find poles in other partial waves beyond the $S$-wave.
We perform the pole searches in all the partial waves up to $J=7/2$.
The positions of the extracted poles together with corresponding quantum numbers are identified and listed in Table~\ref{Tab:allpoles}.
These poles could be regarded as the exotic resonances with hidden charm in $S$-, $P$- or even higher partial waves.

\begin{table}[htbp]
    \centering
    \renewcommand\arraystretch{1.3}
    \caption{Pole positions $z_R$ and spin-parities $J^P$ for the states in different partial waves up to $J=7/2$. ``-'' denotes that no pole is found here.
      \label{Tab:allpoles}}
    \begin{tabular}{p{0.7cm}<{\centering}|*3{p{2.4cm}<{\centering}}}
        \hline
        \hline
         & \multicolumn{3}{c}{$z_R$ [MeV]} \\
        \cline{2-4}
        $J^P$ & Fit A & Fit B & Fit C \\
        \hline
        $\frac12^-$ & $4408.1-i181.5$ & $4413.8-i182.4$ & $4407.1-i178.1$ \\
        \hline
        $\frac12^-$ & $4475.7-i143.6$ & $4502.5-i140.7$ & $4476.4-i143.9$ \\
        \hline
        $\frac12^+$ & $4337.2-i76.3$ & $4325.4-i85.4$ & $4335.1-i75.3$ \\
        \hline
        $\frac32^+$ & $4413.0-i197.0$ & $4410.9-i193.0$ & $4413.3-i196.0$ \\
        \hline
        $\frac32^+$ & $4387.9-i156.3$ & - & $4387.4-i156.1$ \\
        \hline
        $\frac32^-$ & $4408.0-i168.9$ & $4408.3-i175.9$ & $4407.0-i167.6$ \\
        \hline
        $\frac32^-$ & $4506.6-i136.2$ & $4501.5-i132.3$ & $4506.5-i133.9$ \\
        \hline
        $\frac52^-$ & $4411.6-i180.1$ & - & $4411.6-i179.9$ \\
        \hline
        $\frac52^+$ & $4400.4-i100.6$ & $4393.3-i101.6$ & $4400.9-i100.5$ \\
        \hline
        $\frac52^+$ & $4467.1-i100.3$ & $4457.8-i140.4$ & $4466.7-i103.2$ \\
        \hline
        $\frac52^+$ & $4439.1-i122.6$ & - & $4437.9-i121.9$ \\
        \hline
        $\frac72^-$ & $4411.5-i179.8$ & - & $4408.2-i182.2$ \\
        \hline
        \hline
    \end{tabular}
\end{table}

It can be seen that the results of Fits~A and C are quite similar, and their poles could be basically regarded as  one-to-one correspondence.
For Fit~B, some poles are not obtained, and the poles listed in the same row may not necessarily be corresponding to the same state.
This indicates that the properties or even the existence of the poles further away from the real axis are uncertain. 
Note that poles with imaginary parts greater than 200~MeV are not considered and listed here.
Nevertheless, some poles in Table~\ref{Tab:allpoles} have relatively stable positions in all three fits, so that our conclusions for such poles are more robust.
Among these poles, the one located at 4335~MeV with $J^P=1/2^+$ is very close to the LHCb $P_c(4337)$ state, but its width is much larger than that from the experiment, which is only about 29~MeV.
The pole at 4387~MeV with $J^P=3/2^+$ is very stable in Fits~A and C, but not observed in Fit~B. Nevertheless, we may associate it with the broad $P_c(4380)$ state proposed in 2015~\cite{LHCb:2015yax}, where its spin was given as $3/2$ or $5/2$ and its parity was not determined. 
Although this $P_c(4380)$ state is not included in the updated experimental result in 2019~\cite{LHCb:2019kea}, its existence has not been excluded either. 
In Ref.~\cite{Ronchen:2012eg,Wang:2022oof}, we also found a state in a similar position, but with quantum number $J^P=5/2^+$. In the current analysis, the position of the $5/2^+$ state has moved to 4400~MeV, which is not a significant change. 

We regard the differences of the pole parameters in the three different fits as a rough estimation of the uncertainties of our results. A more detailed statistical analysis will be possible once more data are available. Moreover, there are systematic uncertainties connected to the theoretical formalism and  improvements can be made in our model in the future.
For example, more channels, including other hidden charm and light decay channels, as well as possible $t$-channel scalar meson exchange processes can be included in the coupled channel calculations. 
Further, in this work, we use SU(3) and SU(4) flavor symmetry to relate all the involved coupling constants as in Ref.~\cite{Wang:2022oof}, instead of using heavy quark spin symmetry (HQSS) as in Refs.~\cite{Du:2019pij,Du:2021fmf}.
Due to the varying degrees of symmetry breaking, how we establish connections between coupling parameters will inevitably have an impact on the results, which can also be regarded as an area for improvement in theory.
%
However, without more experimental data, the significance of the improvements that can be made is still very limited.
In particular, data not only for the $\Lambda_b^0 \to J/\psi p K^-$ decay, but also, e.g. for the $\bar{D} \Lambda_c$ and $\bar{D} \Sigma_c$ mass distributions and other observables would be helpful to further fix the resonance properties. 
A such substantiated data base would strongly reduce the uncertainties of our results, especially those tied to the statistical data analysis.
%

\section{Summary and outlook} \label{Sec:summary}

To summarize our findings, the interactions of $\bar{D}^{(*)} \Lambda_c - \bar{D}^{(*)} \Sigma_c^{(*)} - J/\psi p$ have been studied in the J\"ulich-Bonn dynamical coupled-channel model.
After including the $J/\psi p$ channel, we are able to fit the experimental observables via the two-point loop diagram of the $\Lambda_b^0 \to J/\psi p K^-$ decay.
A series of bound states and resonances with different spin and parity can be dynamically generated in the coupled-channel formalism.
We find two narrow states with $J^P=1/2^-$ and two with $J^P=3/2^-$ associated with the peak structures of the $J/\psi p$ invariant mass distribution. 
Based on their positions, we can associate them with the hidden-charm pentaquark states $P_c$ reported by the LHCb Collaboration.
The other properties of these states, such as the coupling strengths to different channels, are also provided in the analysis.
In addition, we have searched for poles that are not so close to the real axis in the complex planes in different partial waves.
It turns out that the uncertainty of these poles with larger widths, whether regarding its properties or existence, is larger than that of the four narrow $S$-wave poles.
Once there is a significant increase in the data basis for the various processes of the hidden-charm sector in the experiment, such as provided by the Upgrade~II of LHCb, we will be able to obtain a more solid conclusion on these pentaquark states.
%

\section*{Acknowledgments}

The authors gratefully acknowledge computing time on the supercomputer JURECA~\cite{JURECA} at Forschungszentrum J\"ulich under grant no. ``baryonspectro".
This work is supported by the NSFC and the Deutsche Forschungsgemeinschaft (DFG, German Research
Foundation) through the funds provided to the Sino-German Collaborative
Research Center TRR110 “Symmetries and the Emergence of Structure in QCD”
(NSFC Grant No. 12070131001, DFG Project-ID 196253076 - TRR 110), by the NSFC
Grant No.11835015, No.12047503, and  by the Chinese Academy of Sciences (CAS) under Grant No.XDB34030000,
by  the  Chinese  Academy  of Sciences (CAS) President's  International  Fellowship  Initiative
(PIFI)  (grant  no.    2018DM0034). The work of UGM and DR is further supported  by the Deutsche Forschungsgemeinschaft (DFG,
German Research Foundation) as part of the CRC 1639 NuMeriQS – project
no. 511713970. Also funded by the Deutsche Forschungsgemeinschaft (DFG, German Research Foundation) – 491111487.
\begin{appendix}

\section{Cut-off values} \label{App:cutoff}

As with the applied strategy in previous work, we take one cut-off for each process, rather than for each vertex. 
For the cut-offs in $u$-channel diagrams, whether the exchanged particle is charmed baryon $\Lambda_c$, $\Sigma_c$, or doubly charmed baryon $\Xi_{cc}$, we always fix their values to 3~GeV as previously.
The contribution of the $u$-channel diagrams is very small, and this value can hardly affect our results.
In Table~\ref{Tab:sum}, we list all the $t$-channel processes with the exchanged particle considered in the present work, and their corresponding cut-off values in all three sets of fit results are also provided.
And the specific expressions of different types of potentials can be found in the Appendix of Ref.~\cite{Wang:2022oof}.

\begin{table}[htbp]
    \centering
    \renewcommand\arraystretch{1.25}
    \caption{The values of cut-offs in form factors for all the considered $t$-channel processes. ``Ex'' is the exchanged particle. ``Ty'' denotes the types of potentials as defined in the Appendix of Ref.~\cite{Wang:2022oof}. ``*'' indicates that it is not a free parameter in the fit. \label{Tab:sum}}
    \begin{tabular}{p{2.5cm}<{\centering}|p{0.7cm}<{\centering}|p{0.6cm}<{\centering}|*3{p{1.3cm}<{\centering}}}
        \hline
        \hline
         & & & \multicolumn{3}{c}{$\Lambda$ [MeV]} \\
        \cline{4-6}
        Process & Ex & Ty & Fit A & Fit B & Fit C \\
        \hline
        $\bar{D} \Lambda_c \to \bar{D} \Lambda_c$ & $\omega$ & 1 & 1459 & 2528 & 834 \\
        $\bar{D} \Lambda_c \to \bar{D} \Sigma_c$ & $\rho$ & 1 & 2893 & 2926 & 2893 \\
        $\bar{D} \Lambda_c \to \bar{D}^* \Lambda_c$ & $\omega$ & 4 & 2990 & 2943 & 2868 \\
        $\bar{D} \Lambda_c \to \bar{D}^* \Sigma_c$ & $\pi$ & 3 & 800* & 800* & 800* \\
         & $\rho$ & 4 & 2149 & 2200 & 2149 \\
        $\bar{D} \Lambda_c \to J/\psi N$ & $D$ & 3 & 2000* & 2000* & 2000* \\
         & $D^*$ & 4 & 2000* & 2000* & 2000* \\
        $\bar{D} \Lambda_c \to \bar{D} \Sigma_c^*$ & $\rho$ & 6 & 1877 & 1718 & 1877 \\
        $\bar{D} \Sigma_c \to \bar{D} \Sigma_c$ & $\rho$ & 1 & 2818 & 2654 & 2823 \\
         & $\omega$ & 1 & 2990 & 2754 & 3000 \\
        $\bar{D} \Sigma_c \to \bar{D}^* \Lambda_c$ & $\pi$ & 3 & 800* & 800* & 800* \\
         & $\rho$ & 4 & 1634 & 1704 & 1634 \\
        $\bar{D} \Sigma_c \to \bar{D}^* \Sigma_c$ & $\pi$ & 3 & 800* & 800* & 800* \\
         & $\eta$ & 3 & 1500* & 3000* & 1508 \\
         & $\eta^\prime$ & 3 & 1500* & 3000* & 1150 \\
         & $\rho$ & 4 & 1660 & 1680 & 1660 \\
         & $\omega$ & 4 & 1375 & 1432 & 1369 \\
        $\bar{D} \Sigma_c \to J/\psi N$ & $D$ & 3 & 2000* & 2000* & 2000* \\
         & $D^*$ & 4 & 2000* & 2000* & 2000* \\
        $\bar{D} \Sigma_c \to \bar{D} \Sigma_c^*$ & $\rho$ & 6 & 1636 & 1560 & 1636 \\
         & $\omega$ & 6 & 981 & 913 & 988 \\
        $\bar{D}^* \Lambda_c \to \bar{D}^* \Lambda_c$ & $\omega$ & 8 & 2933 & 2783 & 2960 \\
        $\bar{D}^* \Lambda_c \to \bar{D}^* \Sigma_c$ & $\pi$ & 9 & 800* & 800* & 800* \\
         & $\rho$ & 8 & 2452 & 2193 & 2452 \\
        $\bar{D}^* \Lambda_c \to J/\psi N$ & $D$ & 9 & 2000* & 2000* & 2000* \\
         & $D^*$ & 8 & 2000* & 2000* & 2000* \\
        $\bar{D}^* \Lambda_c \to \bar{D} \Sigma_c^*$ & $\pi$ & 11 & 800* & 800* & 800* \\
         & $\rho$ & 12 & 1454 & 1506 & 1454 \\
        $\bar{D}^* \Sigma_c \to \bar{D}^* \Sigma_c$ & $\pi$ & 9 & 800* & 800* & 843 \\
         & $\eta$ & 9 & 3000* & 3000* & 1389 \\
         & $\eta^\prime$ & 9 & 3000* & 3000* & 1637 \\
         & $\rho$ & 8 & 2600 & 2577 & 2515 \\
         & $\omega$ & 8 & 800 & 876 & 1163 \\
        $\bar{D}^* \Sigma_c \to J/\psi N$ & $D$ & 9 & 2000* & 2000* & 2000* \\
         & $D^*$ & 8 & 2000* & 2000* & 2000* \\
        $\bar{D}^* \Sigma_c \to \bar{D} \Sigma_c^*$ & $\pi$ & 11 & 800* & 800* & 800* \\
         & $\eta$ & 11 & 3000* & 800* & 3000 \\
         & $\eta^\prime$ & 11 & 3000* & 800* & 2359 \\
         & $\rho$ & 12 & 1669 & 1705 & 1669 \\
         & $\omega$ & 12 & 1780 & 1870 & 1787 \\
        $J/\psi N \to \bar{D} \Sigma_c^*$ & $D$ & 11 & 2000* & 2000* & 2000* \\
         & $D^*$ & 12 & 2000* & 2000* & 2000* \\
        $\bar{D} \Sigma_c^* \to \bar{D} \Sigma_c^*$ & $\rho$ & 14 & 1664 & 1645 & 1669 \\
         & $\omega$ & 14 & 1597 & 1755 & 1593 \\
        \hline
        \hline
    \end{tabular}
\end{table}

\end{appendix}

\bibliographystyle{plain}

\begin{thebibliography}{99}

\bibitem{LHCb:2015yax}
R.~Aaij \textit{et al.} [LHCb],
Phys. Rev. Lett. \textbf{115}, 072001 (2015).

\bibitem{LHCb:2019kea}
R.~Aaij \textit{et al.} [LHCb],
Phys. Rev. Lett. \textbf{122}, no.22, 222001 (2019).

\bibitem{LHCb:2021chn}
R.~Aaij \textit{et al.} [LHCb],
Phys. Rev. Lett. \textbf{128}, no.6, 062001 (2022).

\bibitem{Johnson:2024omq}
D.~Johnson, I.~Polyakov, T.~Skwarnicki and M.~Wang,
[arXiv:2403.04051 [hep-ex]].

\bibitem{Wu:2010jy}
J.~J.~Wu, R.~Molina, E.~Oset and B.~S.~Zou,
Phys. Rev. Lett. \textbf{105}, 232001 (2010) ;
Phys. Rev. C \textbf{84}, 015202 (2011).

\bibitem{Wang:2011rga}
W.~L.~Wang, F.~Huang, Z.~Y.~Zhang and B.~S.~Zou,
Phys. Rev. C \textbf{84}, 015203 (2011).

\bibitem{Yang:2011wz}
Z.~C.~Yang, Z.~F.~Sun, J.~He, X.~Liu and S.~L.~Zhu,
Chin. Phys. C \textbf{36}, 6-13 (2012).

\bibitem{Yuan:2012wz}
S.~G.~Yuan, K.~W.~Wei, J.~He, H.~S.~Xu and B.~S.~Zou,
Eur. Phys. J. A \textbf{48}, 61 (2012).

\bibitem{Wu:2012md}
J.~J.~Wu, T.~S.~H.~Lee and B.~S.~Zou,
Phys. Rev. C \textbf{85}, 044002 (2012).

\bibitem{Xiao:2013yca}
C.~W.~Xiao, J.~Nieves and E.~Oset,
Phys. Rev. D \textbf{88}, 056012 (2013).

\bibitem{Uchino:2015uha}
T.~Uchino, W.~H.~Liang and E.~Oset,
Eur. Phys. J. A \textbf{52}, no.3, 43 (2016).

\bibitem{Chen:2015loa}
R.~Chen, X.~Liu, X.~Q.~Li and S.~L.~Zhu,
Phys. Rev. Lett. \textbf{115}, no.13, 132002 (2015).

\bibitem{Chen:2015moa}
H.~X.~Chen, W.~Chen, X.~Liu, T.~G.~Steele and S.~L.~Zhu,
Phys. Rev. Lett. \textbf{115}, no.17, 172001 (2015).

\bibitem{He:2015cea}
J.~He,
Phys. Lett. B \textbf{753}, 547-551 (2016).

\bibitem{Burns:2015dwa}
T.~J.~Burns,
Eur. Phys. J. A \textbf{51}, no.11, 152 (2015).

\bibitem{Huang:2015uda}
H.~Huang, C.~Deng, J.~Ping and F.~Wang,
Eur. Phys. J. C \textbf{76}, no.11, 624 (2016).

\bibitem{Chen:2016heh}
R.~Chen, X.~Liu and S.~L.~Zhu,
Nucl. Phys. A \textbf{954}, 406-421 (2016).

\bibitem{Roca:2016tdh}
L.~Roca and E.~Oset,
Eur. Phys. J. C \textbf{76}, no.11, 591 (2016).

\bibitem{Lu:2016nnt}
Q.~F.~L\"u and Y.~B.~Dong,
Phys. Rev. D \textbf{93}, no.7, 074020 (2016).

\bibitem{Shen:2016tzq}
C.~W.~Shen, F.~K.~Guo, J.~J.~Xie and B.~S.~Zou,
Nucl. Phys. A \textbf{954}, 393-405 (2016).

\bibitem{Ortega:2016syt}
P.~G.~Ortega, D.~R.~Entem and F.~Fern\'andez,
Phys. Lett. B \textbf{764}, 207-211 (2017).

\bibitem{Lin:2017mtz}
Y.~H.~Lin, C.~W.~Shen, F.~K.~Guo and B.~S.~Zou,
Phys. Rev. D \textbf{95}, no.11, 114017 (2017).

\bibitem{Liu:2019tjn}
M.~Z.~Liu, Y.~W.~Pan, F.~Z.~Peng, M.~S\'anchez S\'anchez, L.~S.~Geng, A.~Hosaka and M.~Pavon Valderrama,
Phys. Rev. Lett. \textbf{122}, no.24, 242001 (2019).

\bibitem{Guo:2019kdc}
Z.~H.~Guo and J.~A.~Oller,
Phys. Lett. B \textbf{793}, 144-149 (2019).

\bibitem{Liu:2019zvb}
M.~Z.~Liu, T.~W.~Wu, M.~S\'anchez S\'anchez, M.~P.~Valderrama, L.~S.~Geng and J.~J.~Xie,
Phys. Rev. D \textbf{103}, no.5, 054004 (2021).

\bibitem{Du:2019pij}
M.~L.~Du, V.~Baru, F.~K.~Guo, C.~Hanhart, U.-G.~Mei\ss{}ner, J.~A.~Oller and Q.~Wang,
Phys. Rev. Lett. \textbf{124}, no.7, 072001 (2020).

\bibitem{Xiao:2020frg}
C.~W.~Xiao, J.~X.~Lu, J.~J.~Wu and L.~S.~Geng,
Phys. Rev. D \textbf{102}, no.5, 056018 (2020).

\bibitem{Du:2021fmf}
M.~L.~Du, V.~Baru, F.~K.~Guo, C.~Hanhart, U.-G.~Mei\ss{}ner, J.~A.~Oller and Q.~Wang,
JHEP \textbf{08}, 157 (2021).

\bibitem{Nakamura:2021qvy}
S.~X.~Nakamura,
Phys. Rev. D \textbf{103}, 111503 (2021).

\bibitem{Guo:2015umn}
F.~K.~Guo, U.-G.~Mei\ss{}ner, W.~Wang and Z.~Yang,
Phys. Rev. D \textbf{92}, no.7, 071502 (2015).

\bibitem{Burns:2022uiv}
T.~J.~Burns and E.~S.~Swanson,
Phys. Rev. D \textbf{106}, no.5, 054029 (2022).

\bibitem{Maiani:2015vwa}
L.~Maiani, A.~D.~Polosa and V.~Riquer,
Phys. Lett. B \textbf{749}, 289-291 (2015).

\bibitem{Li:2015gta}
G.~N.~Li, X.~G.~He and M.~He,
JHEP \textbf{12}, 128 (2015).

\bibitem{Ghosh:2015xqp}
R.~Ghosh, A.~Bhattacharya and B.~Chakrabarti,
Phys. Part. Nucl. Lett. \textbf{14}, no.4, 550-552 (2017).

\bibitem{Wang:2015epa}
Z.~G.~Wang,
Eur. Phys. J. C \textbf{76}, no.2, 70 (2016).

\bibitem{Xiang:2017byz}
J.~B.~Xiang, H.~X.~Chen, W.~Chen, X.~B.~Li, X.~Q.~Yao and S.~L.~Zhu,
Chin. Phys. C \textbf{43}, no.3, 034104 (2019).

\bibitem{Hiyama:2018ukv}
E.~Hiyama, A.~Hosaka, M.~Oka and J.~M.~Richard,
Phys. Rev. C \textbf{98}, no.4, 045208 (2018).

\bibitem{Ali:2019npk}
A.~Ali and A.~Y.~Parkhomenko,
Phys. Lett. B \textbf{793}, 365-371 (2019).


\bibitem{Liu:2015fea}
X.~H.~Liu, Q.~Wang and Q.~Zhao,
Phys. Lett. B \textbf{757}, 231-236 (2016).

\bibitem{Guo:2016bkl}
F.~K.~Guo, U.-G.~Mei\ss{}ner, J.~Nieves and Z.~Yang,
Eur. Phys. J. A \textbf{52}, no.10, 318 (2016).

\bibitem{Roca:2015dva}
L.~Roca, J.~Nieves and E.~Oset,
Phys. Rev. D \textbf{92}, no.9, 094003 (2015).

\bibitem{Shimizu:2016rrd}
Y.~Shimizu, D.~Suenaga and M.~Harada,
Phys. Rev. D \textbf{93}, no.11, 114003 (2016).

\bibitem{Yamaguchi:2016ote}
Y.~Yamaguchi and E.~Santopinto,
Phys. Rev. D \textbf{96}, no.1, 014018 (2017).

\bibitem{Shimizu:2017xrg}
Y.~Shimizu and M.~Harada,
Phys. Rev. D \textbf{96}, no.9, 094012 (2017).

\bibitem{Yamaguchi:2017zmn}
Y.~Yamaguchi, A.~Giachino, A.~Hosaka, E.~Santopinto, S.~Takeuchi and M.~Takizawa,
Phys. Rev. D \textbf{96}, no.11, 114031 (2017).

\bibitem{Xiao:2019aya}
C.~W.~Xiao, J.~Nieves and E.~Oset,
Phys. Rev. D \textbf{100}, no.1, 014021 (2019).

\bibitem{Burns:2019iih}
T.~J.~Burns and E.~S.~Swanson,
Phys. Rev. D \textbf{100}, no.11, 114033 (2019).

\bibitem{Yalikun:2021bfm}
N.~Yalikun, Y.~H.~Lin, F.~K.~Guo, Y.~Kamiya and B.~S.~Zou,
Phys. Rev. D \textbf{104}, no.9, 094039 (2021).


\bibitem{Schutz:1998jx}
C.~Sch\"utz, J.~Haidenbauer, J.~Speth and J.~W.~Durso,
Phys. Rev. C \textbf{57}, 1464-1477 (1998).



\bibitem{Ronchen:2012eg}
D.~R\"onchen, M.~D\"oring, F.~Huang, H.~Haberzettl, J.~Haidenbauer, C.~Hanhart, S.~Krewald, U.-G.~Mei\ss{}ner and K.~Nakayama,
Eur. Phys. J. A \textbf{49}, 44 (2013).

\bibitem{Ronchen:2014cna}
D.~R\"onchen, M.~D\"oring, F.~Huang, H.~Haberzettl, J.~Haidenbauer, C.~Hanhart, S.~Krewald, U.-G.~Mei\ss{}ner and K.~Nakayama,
Eur. Phys. J. A \textbf{50}, no.6, 101 (2014),
[erratum: Eur. Phys. J. A \textbf{51}, no.5, 63 (2015)].


\bibitem{Ronchen:2022hqk}
D.~R\"onchen, M.~D\"oring, U.-G.~Mei\ss{}ner and C.~W.~Shen,
Eur. Phys. J. A \textbf{58}, no.11, 229 (2022).

\bibitem{Wang:2022osj}
Y.~F.~Wang, D.~R\"onchen, U.-G.~Mei\ss{}ner, Y.~Lu, C.~W.~Shen and J.~J.~Wu,
Phys. Rev. D \textbf{106}, no.9, 094031 (2022).

\bibitem{Mai:2021vsw}
M.~Mai \textit{et al.} [J\"ulich-Bonn-Washington],
Phys. Rev. C \textbf{103}, no.6, 065204 (2021).

\bibitem{Mai:2021aui}
M.~Mai \textit{et al.} [J\"ulich-Bonn-Washington],
Phys. Rev. C \textbf{106}, no.1, 015201 (2022).

\bibitem{Mai:2023cbp}
M.~Mai \textit{et al.} [J\"ulich\textendash{}Bonn\textendash{}Washington],
Eur. Phys. J. A \textbf{59}, no.12, 286 (2023).

\bibitem{Shen:2017ayv}
C.~W.~Shen, D.~R\"onchen, U.-G.~Mei\ss{}ner and B.~S.~Zou,
Chin. Phys. C \textbf{42}, no.2, 023106 (2018).

\bibitem{Wang:2022oof}
Z.~L.~Wang, C.~W.~Shen, D.~R\"onchen, U.~G.~Mei\ss{}ner and B.~S.~Zou,
Eur. Phys. J. C \textbf{82}, no.5, 497 (2022).

\bibitem{GlueX:2019mkq}
A.~Ali \textit{et al.} [GlueX],
Phys. Rev. Lett. \textbf{123}, no.7, 072001 (2019).

\bibitem{Goncalves:2019vvo}
V.~P.~Gon\c{c}alves and M.~M.~Jaime,
Phys. Lett. B \textbf{805}, 135447 (2020).

\bibitem{Wang:2023thy}
X.~Wang, X.~Cao, A.~Guo, L.~Gong, X.~S.~Kang, Y.~T.~Liang, J.~J.~Wu and Y.~P.~Xie,
[arXiv:2311.07008 [hep-ph]].

\bibitem{Doring:2009yv}
M.~D\"oring, C.~Hanhart, F.~Huang, S.~Krewald and U.-G.~Mei\ss{}ner,
Nucl. Phys. A \textbf{829}, 170-209 (2009).

\bibitem{Gasparyan:PhDthesis}
A.~Gasparyan,
Ph.D. thesis, Universit\"at Bonn, Bonn, Germany, 2003.

\bibitem{Xiao:2016ogq}
C.~W.~Xiao,
Phys. Rev. D \textbf{95}, no.1, 014006 (2017).

\bibitem{Wu:2009tu}
J.~J.~Wu, S.~Dulat and B.~S.~Zou,
Phys. Rev. D \textbf{80}, 017503 (2009).

\bibitem{JURECA}
Jülich Supercomputing Centre,  
Journal of large-scale research facilities, 7, A182 (2021). 

\bibitem{Wang:2023snv}
Y.~F.~Wang, U.-G.~Mei\ss{}ner, D.~R\"onchen and C.~W.~Shen,
Phys. Rev. C \textbf{109}, no.1, 015202 (2024).
\end{thebibliography}

\end{document}